\documentclass[twocolumn]{aastex63}
\usepackage{amsmath}
\usepackage{amssymb}
\usepackage{graphicx}
\usepackage[caption=false]{subfig}

\newcommand{\cha}{Chandra}
\newcommand{\Msun}{${\rm M}_{\odot}$}
\newcommand{\maxiJ}{MAXI\,J1820$+$070}
\newcommand{\sw}{Swift}
\newcommand{\lx}{$L_{\rm X}$}
\newcommand{\lr}{$L_{\rm R}$}
\newcommand{\ledd}{$L_{\rm Edd}$}

\received{September 8, 2020}
\revised{November 24, 2020}
\accepted{December 7, 2020}

\submitjournal{ApJ}

\shorttitle{Disk/jet coupling of MAXI\,J1820+070}
\shortauthors{A. W. Shaw et al.}

\graphicspath{{./}{Figures/}}

\begin{document}

\title{Observations of the Disk/Jet Coupling of \maxiJ\ During its Descent to Quiescence}

\correspondingauthor{A. W. Shaw}
\email{aarrans@unr.edu}

\author[0000-0002-8808-520X]{A. W. Shaw}
\affiliation{Department of Physics, University of Nevada, Reno, NV 89557, USA}

\author[0000-0002-7092-0326]{R. M. Plotkin}
\affiliation{Department of Physics, University of Nevada, Reno, NV 89557, USA}

\author[0000-0003-3124-2814]{J. C. A. Miller-Jones}
\affiliation{International Centre for Radio Astronomy Research, Curtin University, GPO Box U1987, Perth, WA 6845, Australia}

\author[0000-0001-8371-2713]{J. Homan}
\affiliation{Eureka Scientific, Inc., 2452 Delmer Street, Oakland, CA 94602, USA}
\affiliation{SRON Netherlands Institute for Space Research, Sorbonnelaan 2, NL-3584 CA Utrecht, Netherlands}

\author[0000-0001-5802-6041]{E. Gallo}
\affiliation{Department of Astronomy, University of Michigan, 1085 S University, Ann Arbor, MI 48109, USA}

\author[0000-0002-3500-631X]{D. M. Russell}
\affiliation{Center for Astro, Particle and Planetary Physics, New York University Abu Dhabi, PO Box 129188, Abu Dhabi, UAE}

\author[0000-0001-5506-9855]{J. A. Tomsick}
\affiliation{Space Sciences Laboratory, 7 Gauss Way, University of California, Berkeley, CA 94720-7450, USA}

\author[0000-0002-3638-0637]{P. Kaaret}
\affiliation{Department of Physics and Astronomy, University of Iowa, Iowa City, IA 52242, USA}

\author[0000-0001-5538-5831]{S. Corbel}
\affiliation{AIM, CEA, CNRS,  Université de Paris, Université Paris-Saclay, F-91191 Gif-sur-Yvette, France}
\affiliation{Station de Radioastronomie de Nançay, Observatoire de Paris, PSL Research University, CNRS, Univ. Orléans, 18330 Nançay, France}

\author[0000-0001-9075-1489]{M. Espinasse}
\affiliation{AIM, CEA, CNRS,  Université de Paris, Université Paris-Saclay, F-91191 Gif-sur-Yvette, France}

\author[0000-0002-7735-5796]{J. Bright}
\affiliation{Astrophysics, Department of Physics, University of Oxford, Keble Road, Oxford, OX1 3RH, UK}

\begin{abstract}

Black hole X-ray binaries in the quiescent state (Eddington ratios typically $\lesssim$10$^{-5}$) display softer X-ray spectra (photon indices $\Gamma\sim2$) compared to higher-luminosity black hole X-ray binaries in the hard state ($\Gamma\sim1.7$). However, the cause of this softening, and its implications for the underlying accretion flow, are still uncertain. Here, we present quasi-simultaneous X-ray and radio spectral monitoring of the black hole X-ray binary \maxiJ\ during the decay of its 2018 outburst and of a subsequent re-flare in 2019, providing an opportunity to monitor a black hole X-ray binary as it actively transitions into quiescence. We probe 1--10 keV X-ray luminosities as low as \lx$\sim4\times10^{32}$ erg s$^{-1}$, equivalent to Eddington fractions of $\sim4\times10^{-7}$. During its decay towards quiescence, the X-ray spectrum of \maxiJ\ softens from $\Gamma\sim1.7$ to $\Gamma\sim2$, with the softening taking $\sim30$d, and completing at \lx$\approx10^{34}$ erg s$^{-1}$ ($\approx10^{-5}$ \ledd). While the X-ray spectrum softens, the radio spectrum generally remains flat/inverted throughout the decay. We also find that \maxiJ\ follows a radio (\lr) --- X-ray luminosity correlation of the form \lr$\propto$\lx$^{0.52\pm0.07}$, making it the fourth black hole system to follow the so-called `standard track' unbroken over several (in this case, four) decades in \lx. Comparing the radio/X-ray spectral evolution(s) with the \lr\ --- \lx\ plane, we find that the X-ray softening is consistent with X-rays produced by Comptonization processes in a radiatively inefficient accretion flow. We generally disfavor X-ray emission originating solely from within the jet, with the possible exception of X-rays produced via synchrotron self-Compton processes.

\end{abstract}


\section{Introduction} \label{sec:intro}

Black hole low-mass X-ray binaries (BH-LMXBs) are binary systems containing a black hole (BH) accreting matter from a low mass ($\lesssim1-2$\Msun) stellar companion. Most BH-LMXBs are transients, spending long periods of time in a quiescent state, exhibiting very low accretion rates, followed by bright outbursts during which the source luminosity increases by several orders of magnitude \citep[see e.g.][]{Remillard-2006,TetarenkoB-2016} across the majority of the electromagnetic spectrum. During these outbursts, BH-LMXBs typically transition through a series of `accretion states,' often spending several months in a soft X-ray spectral state characterized by disk-dominated soft X-ray spectra and little to no detectable radio emission. A transition to a hard X-ray state is marked with the emergence of persistent compact radio emission with a flat or inverted spectrum often extending to the infrared regime \citep[e.g.][]{Fender-2001a,Corbel-2002a,Russell-2013a}, interpreted as emission from a partially self-absorbed synchrotron jet. The compact jet may become increasingly important as BH-LMXBs fade to lower luminosities, where the jet’s mechanical power might account for a substantial fraction of the total accretion power \citep{Fender-2001b,Fender-2003b}. BH-LMXBs in the hard state exhibit X-ray spectra consistent with inverse-Compton scattering of photons % off of a `corona' 
off hot electrons, manifesting as a hard power-law with a high-energy cutoff \citep[see e.g.][for reviews of accretion states]{Remillard-2006,Done-2007,Belloni-2010}.

In the quiescent state, BH-LMXBs are observed to have softer X-ray spectra compared to the hard state \citep[e.g.][]{Tomsick-2001,Kong-2002c,Tomsick-2004,Corbel-2006,Reynolds-2014}, with the spectral softening probably occurring at X-ray luminosity \lx $\lesssim10^{-4}$ \ledd\ and typically completing by \lx $\sim10^{-5}$ \ledd, where \ledd\ is the Eddington luminosity, at which point the spectral shape remains constant as the luminosity continues to decrease \citep{Sobolewska-2011,Plotkin-2013,Plotkin-2017a}. However, the cause of the softening remains unknown, as observations with the sensitivities required to accurately measure the spectrum of BH-LMXBs as they approach quiescent luminosities are rare.

At the lowest luminosities, it is generally accepted that most (but not necessarily all) of the X-ray emission is produced by a radiatively inefficient mechanism, for both accretion flow and jet-related origins \citep[e.g.][]{Blandford-1999,Markoff-2003,Markoff-2005c,Narayan-2008,Yuan-2014}. There are a number of varieties of radiatively-inefficient accretion flow (RIAF) models which generally predict that, as luminosity decreases, there will be a gradual increase of the X-ray power law photon index ($\Gamma$). The softening is generally due to either a lower optical depth to inverse Compton scattering, and/or to a lower average energy change per inverse Compton scatter \citep{Esin-1997}. 

Thus, in the context of a RIAF, as a BH-LMXB fades from the hard state into quiescence there should be fewer hard X-rays emitted  \citep[see also][]{Veledina-2011}.\footnote{Note, at the very lowest accretion rates, one expects the X-ray spectrum to be dominated by bremsstrahlung radiation in most RIAF models \citep[e.g.,][]{Yuan-2014}.}

On the other hand, it is also possible for synchrotron and/or synchrotron self-Compton (SSC) emission from a jet to contribute to the X-ray emission \citep[e.g.][]{Yuan-2005,Markoff-2005c,Corbel-2008,Plotkin-2012a}, and there are several scenarios where jetted X-ray emission could explain the X-ray softening between the hard state and quiescence. If X-rays are produced by optically thin synchrotron radiation from non-thermal particles, then from most diffusive shock acceleration scenarios \citep[e.g.][]{Jones-1991} one would expect X-ray photon indices from $\Gamma \approx 1.5-1.7$ in the hard state. Then, the X-ray softening in quiescence could be explained by increased radiative losses, since a synchrotron cooled jet would produce a steeper X-ray spectrum \citep{Heinz-2004}. Alternatively, if radiative losses are never significant, then less efficient particle acceleration could be invoked (i.e., the synchrotron emitting particle distribution steepens with decreasing luminosity), or SSC processes from particles accelerated along the jet and/or thermal particles in the base of the jet. In the SSC case, a spectral softening could occur if a softer spectrum of seed photons get upscattered into the X-ray waveband \citep[see, e.g.,][for discussions on particle acceleration and SSC]{Corbel-2008,Plotkin-2015a,Plotkin-2017a,Connors-2017}.

Although models that suggest jet emission may contribute some level of X-ray emission have shown promise \citep[e.g.][]{Markoff-2001a,Markoff-2003,Plotkin-2015a,Connors-2017}, this is still a matter of debate \citep[e.g.][]{Zdziarski-2003,Maccarone-2005,Malzac-2009}. Progress in clarifying the dominant mechanism driving the softening has largely been hampered by lack of sufficient observations while the X-ray spectrum is actively softening. Part of the reason is because the timing of such observations is difficult. Also, the low-luminosity nature of BH-LMXBs during the transition to quiescence  prevents precise measures of the X-ray spectral shape for all but the closest BH-LMXBs with low line of sight absorption. Furthermore, from X-ray observations alone, it is extremely challenging to differentiate between the scenarios described above. 

One can make meaningful progress, however, by performing high signal-to-noise X-ray spectral monitoring in coordination with radio observations. As hard state BH-LMXBs transition towards quiescence they trace out distinct paths through the \lr --- \lx\ plane, where \lr\ is the radio luminosity, \citep[e.g.][]{Corbel-2013a,Gallo-2014b,Gallo-2018}. The presence of correlated radio and X-ray emission suggests a link between the innermost regions of the accretion flow/jet (probed by X-rays) and the outer regions of the jet (probed by radio observations). So far, three BH-LMXBs have been seen to follow a similar non-linear correlation of the form \lr\ $\propto$ \lx$^{0.5-0.7}$ \citep{Gallo-2014b}, which is often referred to as the `standard track.' In addition, the only three highly sub-Eddington (\lx\ $\sim10^{-8.5}$ \ledd) BH systems with meaningful radio constraints lie on the extrapolation of the standard track \citep{Gallo-2006,Corbel-2013a,Gallo-2014b,Dzib-2015,Ribo-2017,Tremou-2020}. However, a large population of hard state BH-LMXBs that are `radio-faint' compared to the standard track also exists \citep[see e.g.][]{Corbel-2004,Xue-2007,Gallo-2012}. Under certain assumptions, the slope of the \lr\ --- \lx\ correlation can be used to place constraints on the emission mechanisms at work \citep[see also][]{Markoff-2003,Heinz-2003}, such that combining X-ray (spectroscopy) with radio monitoring has the potential to break model degeneracies \citep[see e.g., high-cadence radio/X-ray monitoring of the 2015 decay of V404 Cygni;][]{Plotkin-2017a}.

In this work, we present radio and X-ray observations of the BH-LMXB \maxiJ\ during the decay of its initial outburst in 2018 and of one of the subsequent re-flares that occurred in 2019. In Section \ref{sec:source} we introduce the source, before detailing our observations and data reduction in Section \ref{sec:obs}. We present and discuss our results in Section \ref{sec:results} and summarize in Section \ref{sec:conclusions}. Throughout this work, we adopt a distance of $d=2.96\pm0.33$ kpc, as determined from trigonometric parallax in the radio waveband by \citet{Atri-2020} using very long baseline interferometry.  

\section{MAXI J1820+070}
\label{sec:source}

\maxiJ\ \citep[also known as ASASSN-18ey;][]{Tucker-2018} was discovered as an optical transient in March 2018 by the All-Sky Automated Survey for SuperNovae \citep[ASAS-SN; see][]{Shappee-2014,Kochanek-2017}. Six days after the initial optical detection, the Monitor of All-Sky X-ray Image \citep[ MAXI;][]{Matsuoka-2009} reported a bright X-ray transient associated with the optical source \citep{Kawamuro-2018,Denisenko-2018}. Follow-up observations were performed across the electromagnetic spectrum \citep[e.g.][]{Baglio-2018,Bright-2018,Uttley-2018} and hinted at a potential BH-LMXB nature for the source. The compact object was dynamically confirmed as a $>5.18\pm0.15$ \Msun\ BH by \citet{Torres-2019b}, later refined to $M_{\rm BH}=8.48^{+0.79}_{-0.72} M_{\odot}$ \citep{Torres-2020}.\footnote{Assuming the binary inclination is equivalent to the inclination of the radio jet \citep[$63\pm3^{\circ}$;][]{Atri-2020}} The binary orbital period, $P_{\rm orb}$, of \maxiJ\ is 16.5h \citep{Torres-2019b}.

The initial outburst lasted for almost a year\footnote{See the MAXI light curve: \href{http://maxi.riken.jp/star_data/J1820+071/J1820+071_00055058g_lc_all.gif}{http://maxi.riken.jp/star\_data/\\J1820+071/J1820+071\_00055058g\_lc\_all.gif}.} and approached quiescence in February 2019 \citep{Russell-2019}. However, the source underwent two  re-flares \citep[see][for the classification scheme of BH-LMXB rebrightening episodes]{Zhang-2019}, which lasted $\sim2$ months each, in March \citep{Ulowetz-2019} and August 2019 \citep{Hambsch-2019}, as well as a third, shorter, re-flare in February 2020 \citep{Adachi-2020}.% before a more permanent return to quiescence.

The brightness of \maxiJ\ \citep[\lx$\sim0.15$ \ledd\ at peak;][]{Atri-2020} during its outburst has made it an exciting candidate for studies of accretion and outflow in BH-LMXBs. For example, `reverberation lags,' lags between the corona and the irradiated accretion disk, showed, for the first time, evidence for a contracting corona in a BH-LMXB \citep{Kara-2019}. During the hard-to-soft state transition of the main outburst, \maxiJ\ launched long-lived bipolar radio and X-ray ejecta that imply jet powers far larger than inferred from radio flares during the soft-state transition \citep{Bright-2020,Homan-2020,Espinasse-2020}. Finally, observations of \maxiJ\ during the soft state revealed an excess emission component in the X-ray spectrum, which has been interpreted as originating in the `plunge' region where matter begins to fall freely into the BH \citep{Fabian-2020}. Contrary to many of the studies discussed above, the work we present here focuses on the source as it approaches quiescence, in an effort to constrain the emission mechanisms at work in BH-LMXBs at low luminosities.

\section{Observations and Analysis}
\label{sec:obs}

Observations were assembled through multiple target of opportunity (ToO) programs designed to piece together dense multi-wavelength spectral monitoring of the entire hard state decay of a transient black hole. For the beginning of the decay (\lx$\approx 10^{-2}$ to $10^{-4}$ \ledd), covering MJD 58397--58519, \maxiJ\ was monitored with the X-ray Telescope (XRT) onboard the Neil Gehrels Swift Observatory \citep{Burrows-2005}. In this work we use data from a combination of guaranteed time observations (GTO; PI Gallo) and regular ToOs, obtaining a total of 21 observations during the decay of the initial outburst, with an average cadence of $\sim2$d. Each GTO Swift epoch was coordinated within $\pm$1 day of a radio observation taken with the Karl G.\ Jansky Very Large Array (VLA) through program 18A-277 (PI Plotkin). We then triggered a joint Chandra/VLA proposal to monitor the source at lower luminosities (Chandra proposal ID 19400238; PI Gallo). However, we were only able to obtain a single epoch on MJD 58441 before the source became Sun-constrained (although we note that we did continue to monitor in the radio during the X-ray Sun constraint, albeit less frequently). Prior to and following the Sun constraint, \maxiJ\ was also observed three times through a joint Chandra/VLA program on MJD 58436, 58518-58519 \citep[Chandra proposal ID 19400337; PI Corbel; see][]{Espinasse-2020}. Finally, after the detection of the first re-flare we triggered another joint Chandra/VLA program to observe its decay six times at low-luminosities (\lx$\lesssim 10^{-4}$ \ledd), typically every $\sim$9d from MJD 58603--58645 (Chandra proposal ID 20400114; PI Gallo). 

For completeness, we also include all Swift/XRT observations available in the archive during the %main outburst decay (MJD=XX -XX) and the 
decay of the first re-flare (covering MJD 58563--58591), and we also include two observations with the X-ray Timing Instrument on the Neutron star Interior Composition Explorer \citep[ NICER/{XTI};][]{Gendreau-2016} close to the end of the main outburst coverage. A summary of all X-ray and radio observations utilized in this work are presented in Tables \ref{tab:X-ray_obs} and \ref{tab:radio_obs} in appendix \ref{sec:obs_tables}.

\subsection{Swift/XRT}

\maxiJ\ was monitored regularly by \sw\ since its initial discovery. In this work we use a subset of the observations spanning the date range 2018 Oct 06 -- 2019 Apr 17 (MJD 58397--58590; Target ID: 10627), covering the decline of the initial outburst of the source and the  decline of the first re-flare. \sw/XRT operated in windowed timing (WT) mode for all of the observations we analyze in this work. All photon counting (PC) mode observations had count rates that were either too high (such that photon pile-up was too great to correct for) or too low (such that not enough photons were collected to construct a good signal-to-noise spectrum).

Data were reprocessed using the {\tt xrtpipeline} tool, part of the {\sc HEAsoft} v6.26.1 software suite for analysis of high energy astrophysical data\footnote{\href{https://heasarc.gsfc.nasa.gov/docs/software/heasoft/}{https://heasarc.gsfc.nasa.gov/docs/software/heasoft/}}. Spectral products were extracted using the {\tt xrtproducts} tool. Source counts were extracted from a circular region 20 pixels ($\approx47\arcsec$) in radius, centered on the source. Background counts were extracted from an annulus centered on the source, with inner and outer radii 80 and 120 pixels, respectively. Response matrices were generated using version 20191017 of the calibration database ({\sc caldb}).

Individual spectra were grouped such that each spectral bin contained a minimum of 15 counts. %, allowing us to use $\chi^2$ as the fit statistic. 
Spectral fits were performed with {\sc xspec} v12.10.1f \citep{Arnaud-1996},  using Cash-statistics modified for background-subtracted spectra \citep[W-statistic;][]{Cash-1979} as the fitting statistic, which is appropriate for fitting low-count spectra whilst also tending to $\chi^2$ in the high-count regime. The majority of spectra were well fit with an absorbed power law model ({\tt powerlaw} in {\sc xspec}), with interstellar absorption accounted for by the {\tt tbabs} model \citep{Wilms-2000}. The first three Swift spectra (MJD 58397--58402) were statistically improved with the addition of a disk blackbody model ({\tt diskbb}) with an inner disk temperature in the range $kT_{\rm in}=0.13$--$0.18$ keV. The presence of the disk component was confirmed by a NICER observation on MJD 58400.9. When including a {\tt diskbb} component for spectral fits at MJD$>$58402, we found the disk normalization was consistent with zero. We extracted unabsorbed 1--10 keV fluxes using the {\tt cflux} model. Uncertainties on the best-fit X-ray parameters are all 90\% confidence unless otherwise stated.

The first 10 Swift observations of our campaign (MJD$<$58420) were taken primarily through our GTO program, where we compensated for the decrease in flux with time by increasing the length of each subsequent Swift/XRT exposure. In turn, we obtained a relatively steady number of X-ray counts in each spectrum. Later in the decay when \maxiJ\ was even fainter, all Swift exposures were relatively short, making it difficult for us to assess degeneracies between best-fit column densities ($N_{\rm H}$) and photon indices ($\Gamma$). Since a primary goal of our program is to quantify changes in X-ray spectral shape, and since we do not expect $N_{\rm H}$ to evolve during the decay, we chose to freeze $N_{\rm H}$ to $1.0\times10^{21}$ cm$^{-2}$ throughout our entire campaign to allow a more uniform comparison between the higher- and lower-count spectra toward the beginning and end of our campaign, respectively. This value is (a) consistent with the Galactic column in the direction of \maxiJ\ \citep{HI4PI-2016} and (b) consistent with the measured values of $N_{\rm H}$ over the course of its outburst \citep[see e.g.][]{Shidatsu-2018,Kajava-2019,Xu-2020}. In addition, during the first 10 Swift observations, where we accounted for the diminishing count rate through increased exposure times, we find a weighted average $N_{\rm H}=1.1\pm0.1\times10^{21}$ cm$^{-2}$ if we allow $N_{\rm H}$ to be free.

\subsection{Chandra/ACIS}

We utilized \cha\ observations of \maxiJ\ over the date range 2018 Nov 13 -- 2019 Jun 11 (MJD 58435--58645), %(PIs: Corbel, Gallo),
which includes observations during the decline of the initial outburst %where \sw\ left off 
as well as the final stages of the decline of the first re-flare. For all observations, the source was placed at the aim point of the S3 chip on the Advanced CCD Imaging Spectrometer \citep[ACIS;][]{Garmire-2003}. Three observations employed the High-Energy Transmission Grating \citep[HETG;][]{Canizares-2005} as a filter in order to avoid photon pile-up. Data were reduced using {\sc ciao} (\cha\ Interactive Analysis of Observations) v4.11 and the \cha\ {\sc caldb} v 4.8.4.1 \citep{Fruscione-2006}. We reprocessed the data using the {\tt chandra\_repro} script to apply the latest calibrations and bad pixel files. Spectral products were extracted using the {\tt specextract} script for the majority of the observations. For each of these, source counts were extracted from a circle of radius $2\arcsec$ centered on the source, whilst we extracted the background from an annular region centered on the source with inner and outer radii $9\arcsec$ and $20\arcsec$, respectively.\footnote{We note that the background region did not include the X-ray jets presented by \citet{Espinasse-2020}} One observation (ObsID 20207) without the HETG in place suffered from extreme photon pileup, for which we were unable to effectively correct. In this case we extracted a spectrum from the CCD read-out streak that appears in bright observations such as this one\footnote{\href{https://cxc.cfa.harvard.edu/ciao/threads/streakextract/}{https://cxc.cfa.harvard.edu/ciao/threads/streakextract/}}. Source counts were extracted using two box regions of size $2\arcsec\times51\arcsec$ centered on each streak on either side of the piled up point source, and background counts from two box regions of size $22\arcsec\times51\arcsec$ at the same location (with the source region excluded). Spectra were extracted with the {\sc ciao} tool {\tt dmextract} and response files created using {\tt mkacisrmf} and {\tt mkarf}. The effective exposure time of the read-out streak spectrum was 296s.

As with the \sw\ data, we used {\sc xspec} to perform spectral fitting. For the shorter \cha\ observations (exposure time $t_{\rm exp}<10$ ks), we chose to bin the spectra such that each bin contained a minimum of one count. For the remaining, longer observations, we grouped the spectra to a minimum of 15 counts per spectral bin, as with the \sw\ spectral fits. All \cha\ spectra were well fit with an absorbed power law model and were not statistically improved with the addition of a {\tt diskbb} model. Unabsorbed 1--10 keV fluxes were extracted in the same way as for the \sw\ data.

\subsection{NICER}

NICER observed \maxiJ\ regularly since its initial discovery. In this work we utilize two observations from 2018 Nov 19 and 21 (MJD 58441 and 58443; ObsIDs 1200120310 and 1200120312; PI Gendreau) covering the initial decline immediately after the \sw\ and \cha\ coverage ended. Both observations were recalibrated with the {\tt nicerl2} task in {\sc HEAsoft} v6.26.1, using {\sc caldb} version 20200202. Spectra were extracted from all active detectors, except detectors \#14 and \#34, which are prone to excessive noise. Background spectra were created using the `3C50\_RGv5' model provided by the NICER team. We used response files recommended for {\sc caldb} version 20200202. 

We followed the same spectral fitting procedure as for the \sw\ data, grouping the data such that each bin contained a minimum of 25 counts and utilizing the \citet{Cash-1979} fitting statistic. Spectra were fit in the 0.5--10 keV range and we included a multiplicative constant of 50/52 in the model to account for the excluded detectors. Both NICER spectra were well fit with an absorbed power law model.

\subsection{VLA}\label{VLAobs}
Our VLA campaign consisted of a total of 19 observations obtained through  NRAO programs 18A-277, SK0335, SJ0238, and SK0114, see Table~\ref{tab:radio_obs}. The first eight VLA observations (MJD 58398--58432) were taken in the  most compact D configuration (maximum baseline ${\rm B}_{\rm max}=1.03$ km), with the next five observations (MJD 58441--58517) in C configuration (${\rm B}_{\rm max}=3.4$ km), and the final six observations (MJD 58603--58645; during the re-flare decay) in B configuration (${\rm B}_{\rm max}=11.1$ km). Observations lasted between $\sim$30--120 min (providing $\sim$1--80 min on source), with longer observations generally toward the end of each decay. 

All data were taken in C-band (4-8 GHz) with 4 GHz total bandwidth ($2\times 2$ GHz basebands centered at 5.0 and 7.0 GHz). For every observation we used scans on 3C\,286  for bandpass calibration and to set the flux density scale (using the \citealt{Perley-2017} coefficients), and we interleaved our science observations with scans on the secondary calibrator J1824+1044 to solve for time-dependent complex gain solutions. Data were reduced using the Common Astronomy Software Application v5.6 \citep[{\tt CASA};][]{McMullin-2007}, and calibrations were performed using the VLA pipeline. A small amount of additional flagging was performed manually prior to imaging the data. 

The data were imaged using {\tt tclean}, using two Taylor terms to model spectral dependences of sources within the field. We used Briggs weighting to reduce sidelobes from other sources in the field, using robust values of 0.0, 0.5, and 1.0 when the array was in D, C, and B configurations, respectively. During the first nine observations \maxiJ\ was $>$1 mJy, and we performed one-to-two rounds of phase-only self-calibration (down to 30-60s solution intervals). We achieved root-mean-square (rms) noise levels ranging from $\approx$3 to $\approx$50 $\mu$Jy bm$^{-1}$ across our entire campaign, as measured from  source-free regions of our radio images. All flux densities were measured at 6 GHz. 

Long-lived ejecta that were launched by \maxiJ\ during its soft state transition (see Section \ref{sec:source}) were present during our coverage of the initial decay.  During our D configuration observations we found it challenging to resolve the core from these ejecta in the image plane, which reached distances up to ~13 arcsec from the compact core. We therefore measured flux densities by fitting the blended core and ejecta in the \textit{uv} plane using {\tt uvmultifit}\footnote{\href{https://github.com/onsala-space-observatory/UVMultiFit}{https://github.com/onsala-space-observatory/UVMultiFit}} \citep{Marti-Vidal-2014}, as described below.

For each D configuration observation we first produced an image in {\tt tclean}, as described above. We then took the sky model produced by {\tt tclean}, masked out the \maxiJ\ complex, and subtracted remaining `field' sources from the visibility set using the task {\tt uvsub}. Depending on the date of observation and the time on source, we expected to detect anywhere from 0-2 relativistic ejecta in each observation in addition to the compact core. We therefore ran three iterations of {\tt uvmultifit}, requiring 1, 2, and 3 point sources.  %providing initial guesses for the positions and flux densities using results from the previous epoch.
For the core we left the radio spectral index $\alpha$ ($f_\nu \propto \nu^\alpha$; where $f_\nu$ is the flux density at frequency $\nu$) as a free parameter, and for the relativistic ejecta we fixed the spectral index to $\alpha=-0.7$ \citep{Bright-2020}. We examined the resulting fit statistics to guide our decision on the best fit (i.e., the 1, 2, or 3 component model); we also examined residual images after subtracting each model (i.e., we used {\tt uvsub} to subtract each model from the visibility set and then created `dirty images' using {\tt tclean}). The peak flux densities and spectral indices of only the core  (from our preferred model) are reported in Table~\ref{tab:radio_obs}. Uncertainties incorporate both rms noise and errors related to the fitting process. Flux densities (and positions) for the relativistic ejecta were reported in \citet{Bright-2020}.

When the VLA was in its C configuration, we could resolve the compact core from the ejecta in the image plane. We generally found for these observations, when the core was fainter, that flux density measurements appeared more reliable in the image plane compare to the \textit{uv} plane.  We therefore measured the peak flux density of the compact core using the task  {\tt imfit} for our C configuration observations, but using a model that required 1, 2, or 3 point-source components. As above, we examined residual images with each model subtracted to decide on the number of required components. The flux densities reported in Table~\ref{tab:radio_obs}  (for our C configuration observations) are the peak flux densities reported by {\tt imfit} only for the core. All of our B configuration observations are during the re-flare decay, by which time radio emission from the relativistic ejecta had completely faded.

For the C and B configuration observations, we measured radio spectral indices by splitting our bandwidth into $4\times 1$ GHz basebands centered at 4.5, 5.5, 6.5 and 7.5 GHz. We then imaged each baseband separately and measured the flux density of the core in {\tt imfit} (using multi-component models in C configuration, as described above). The spectral index was then measured by performing a least-squares-fit, with the uncertainty on $\alpha$ estimated by Monte-Carlo simulations that randomized (and refit) each spectrum \citep[see][]{Plotkin-2017a}. Note, for the D configuration observations, the spectral index was fit as a free parameter when running {\tt uvmultifit}.  Throughout the text, we adopt radio luminosities at 5 GHz %for consistency with previous work, 
as calculated from the measured radio flux density (at 6 GHz), the radio spectral index, and the source distance.

\section{Results and Discussion}
\label{sec:results}

\begin{figure*}
    \centering
        \includegraphics[width=0.8\textwidth,trim={0 0 0 10mm}, clip]{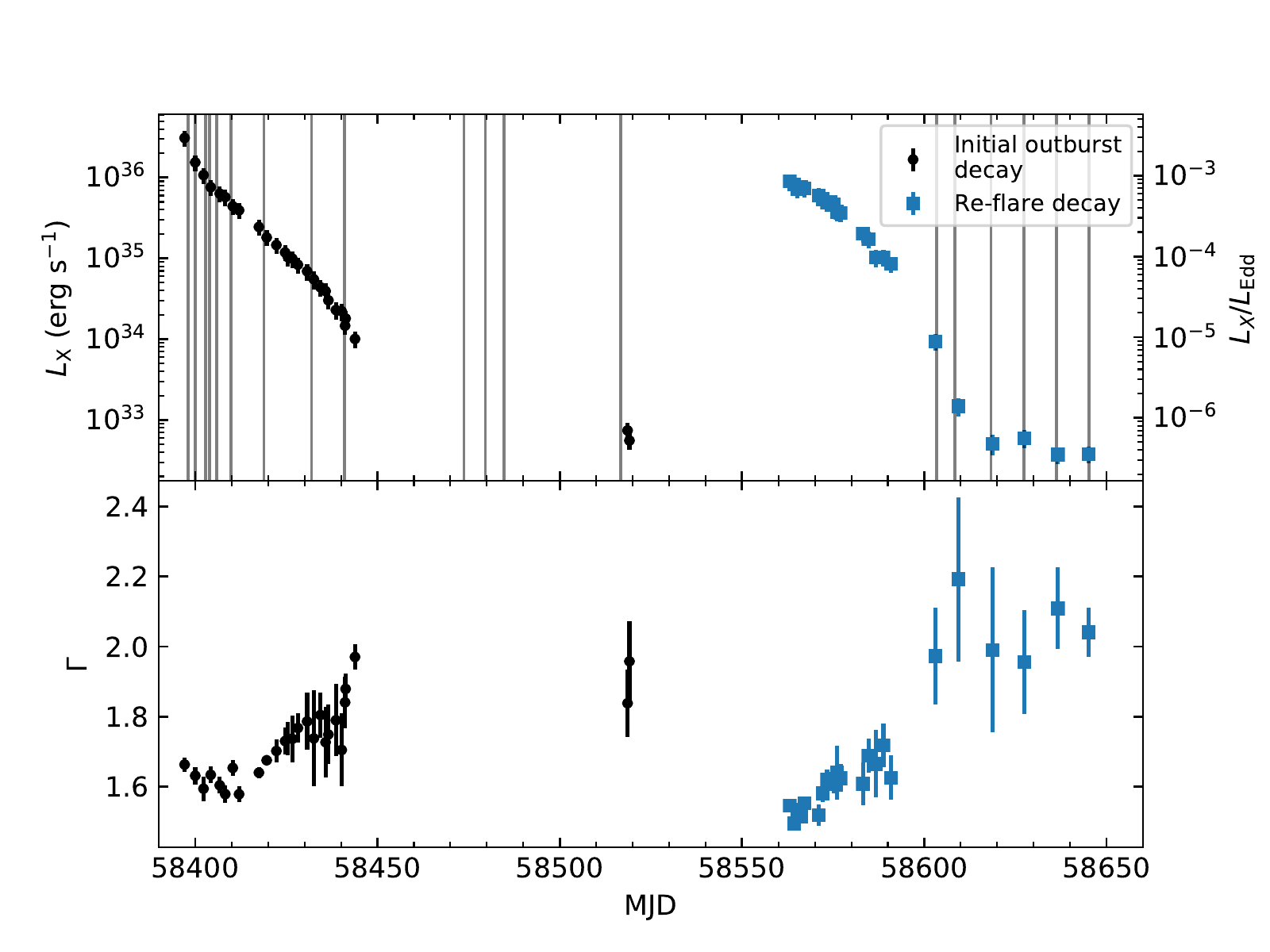}
        \caption{{\em Upper panel:} 1--10 keV X-ray light curve of \maxiJ. The grey vertical lines represent the epochs of the VLA observations. Luminosities were calculated by adopting the distance $d=2.96\pm0.33$ kpc as measured by \citet{Atri-2020}. For the Eddington ratios labeled on the right y-axis, \ledd\ is calculated assuming the mass of the BH $M_{\rm BH}=8.48$ \Msun\ as derived by \citet{Torres-2020}. Plotted \lx\ is unabsorbed. {\em Lower panel:} The best-fit value of the power law index ($\Gamma$) for each observation. In both panels, black, circular points represent observations of the decay of the initial outburst and blue, square points represent observations taken during the decay of the first re-flare.}
    \label{fig:XRlc}%
\end{figure*}

\subsection{X-ray Luminosity and Spectral Evolution}

The X-ray light curve of \maxiJ\ is shown in the upper panel of Figure~\ref{fig:XRlc}, highlighting the end of the initial outburst and the decay of the first re-flare in March 2019. We incorporate the uncertainty on the distance into our luminosity calculation. The decline of both the initial outburst and the first re-flare cover a similar dynamic range in Figure~\ref{fig:XRlc}, with each covering $\sim3$ orders of magnitude in \lx. It is unclear if \maxiJ\ reached its minimum quiescent \lx\ between the end of the initial outburst and the onset of the first re-flare, since X-ray observations of the source at an \lx\ lower than that measured by our final \cha\ epoch do not exist at the time of writing. However, considering the $P_{\rm orb}$ of \maxiJ, we suspect that it may eventually settle to an \lx\ up to 1-2 orders magnitude lower than the minimum observed during our campaign \citep[see, e.g., Figure~4 of][]{Reynolds-2011}.

The lower panel of Figure~\ref{fig:XRlc} shows the evolution of the X-ray spectrum, parameterized by the photon index $\Gamma$. The hard X-ray spectrum ($\Gamma\lesssim1.7$), coupled with the presence of a compact jet, indicates that \maxiJ\ had transitioned to the hard state by the time we commenced our \lr\ --- \lx\ monitoring program. This is confirmed by the detection of significant variability (fractional rms $\gtrsim35$\%) and quasi-periodic oscillations in the \sw\ and NICER light curves \citep{Stiele-2020}, typical of BH-LMXBs in the hard state.

We see a mild hardening of the X-ray spectrum of \maxiJ\ for the first $\approx$15d of our program, with $\Gamma$ evolving from $\approx$1.7 to 1.6 (reaching its hardest values as the source declined from $L_{\rm X}\approx 10^{36.5}$ to $10^{35.6}$ erg s$^{-1}$, equivalent to $\approx 10^{-2.5}$ to $10^{-3.4}$ \ledd. During the first three observations (MJD 58397--58402) we detect a cool, thin disk ($kT=0.13$--$0.18$ keV) in the hard state that slowly fades and is not detectable at later epochs with lower-count spectra (\lx$\lesssim10^{-3}$ \ledd). An accretion disk component is not uncommon in the spectra of BH-LMXBs in the canonical hard state, particularly at these modest Eddington ratios \citep[e.g.][]{Miller-2006b,Reis-2010,Reynolds-2013}. During the decline of the second re-flare of \maxiJ, \citet{Xu-2020} found strong evidence for a truncated disk at \lx$\approx10^{-2.6}$ \ledd, similar to the luminosity at which we require a disk component in the spectra during the decline of the main outburst.

At lower X-ray luminosities during the initial decay, we observe $\Gamma$ evolve from $\sim1.6 $ to $2.0$. The initial hardening and then softening of $\Gamma$ with decreasing X-ray luminosity is a well-known trend \citep[one suggestion is that it is driven by a change in the source of seed photons for inverse Comptonization, see, e.g.,][]{Sobolewska-2011,Kajava-2016}. Based on two Chandra observations taken on MJD 58519, the X-ray spectrum did not appear to continue to soften indefinitely. Unfortunately we cannot empirically confirm this statement, since we lack X-ray coverage between MJDs 58444-58519 because the source was too close to the Sun for X-ray observations. Nevertheless, even though we cannot determine exactly when the X-ray spectrum `saturated' to its maximum $\Gamma$ during the initial decay, we can place a limit that $\Gamma$ saturated on or after MJD 58444, when \lx $\approx 10^{34}$ erg s$^{-1}$ ($\approx10^{-5}$ \ledd). 

Our coverage of the decay of the re-flare started 1--2d after its peak, in contrast with the observations of the decline of the initial outburst, which we commenced $\sim200$d post-peak, after the source had undergone state transitions. During the decay of the re-flare, we only observed the stage of the decay when $\Gamma$ increases. Here, it is more apparent that once $\Gamma \approx 2$ was reached, the X-ray spectrum plateaued to that value, which occurred sometime between the last Swift observation and the first Chandra observation during the re-flare (58591$<$MJD$<$58603, $10^{-4} \gtrsim $\lx/\ledd $\gtrsim 10^{-5}$). The X-ray spectrum then remains soft ($\Gamma\approx2$) over $>1.5$ decades in X-ray luminosity ($10^{-5} \gtrsim $\lx/\ledd $\gtrsim 10^{-6.4}$). 

\subsubsection{Comparisons to Other Quiescent Sources} 
\begin{figure}
    \centering
    \includegraphics[width=0.48\textwidth, trim=5mm 0 5mm 2mm, clip]{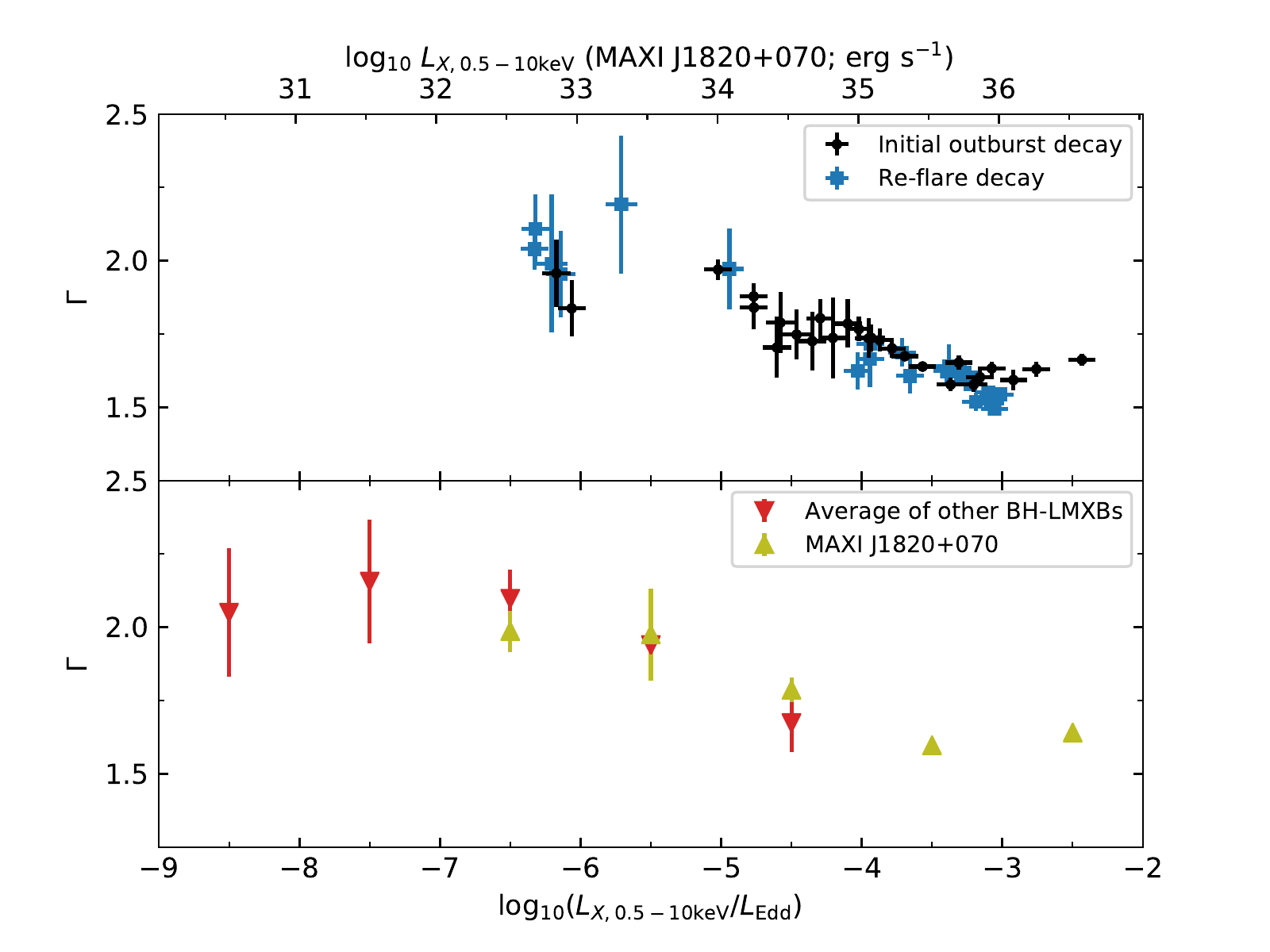}
    \caption{X-ray spectral evolution of \maxiJ\ as a function of Eddington fraction. {\em Upper Panel:} The unbinned data from Figure~\ref{fig:XRlc}, with the same legend. {\em Lower Panel:} The same data, in Eddington fraction bins of width 1 dex (yellow triangles), plotted alongside the photon indices of 10 BH-LMXBs collected by \citet{Plotkin-2013}, with additional data from V404\,Cygni included since its 2015 outburst \citep[][]{Plotkin-2017a}, again in Eddington fraction bins of 1 dex (red inverted triangles). Plotted luminosities for all sources are in the 0.5--10 keV energy band.
    }
    \label{fig:gamma_lum}
\end{figure}

In Figure~\ref{fig:gamma_lum} we show $\Gamma$ as a function of Eddington fraction for both the initial decay and re-flare. This figure demonstrates the softening of the X-ray spectrum (i.e., increasing $\Gamma$) as X-ray luminosity decreases, with $\Gamma$ eventually `saturating' to $\approx 2$ at  \lx $\approx10^{-5}$ \ledd. %\lx $\approx$ $10^{34}$ erg s$^{-1}$ ($\sim10^{-5}$ \ledd). 
The bottom panel of Figure~\ref{fig:gamma_lum} shows the X-ray spectrum binned by X-ray luminosity, to ease comparisons to the ensemble average of other BH-LXMBs in the Chandra archive from \citet{Plotkin-2013}, combined with additional data from V404\,Cygni \citep{Plotkin-2017a}.\footnote{It is also for this ease of comparison that we utilize 0.5--10 keV luminosities of \maxiJ\ in this figure instead of the 1--10 keV luminosities quoted throughout this work.} Amongst all of the quiescent BH-LMXBs in the Chandra archive, the only individual source (besides \maxiJ) where the X-ray softening has been tracked through the `plateau' stage is V404\,Cygni (during its 2015 outburst decay; \citealt{Plotkin-2017a}).\footnote{Though the {\em intermediate}-mass X-ray binary 4U\,1543$-$47 has good coverage of the plateau phase with the {\em Rossi X-ray Timing Explorer} \citep{Kalemci-2005}} For V404\,Cygni, the softening proceeded quickly (on a time-scale of $<$3d) and saturated at a slightly lower luminosity \lx $\approx 10^{-5.5} - 10^{-5.6}$ \ledd.

Comparing \maxiJ, V404\,Cygni, and the ensemble of other quiescent BH-LMXBs, we are starting to uncover potential variations in the X-ray softening between different sources. The general qualitative trend -- that the softening completes below $\approx 10^{-5}$ \ledd\ and then plateaus -- holds. However, the softening appears to occur gradually in \maxiJ, at least according to the initial decay where $\Gamma$ increased over $\sim$30d (we do not have sufficient coverage during the re-flare decay; see Figure~\ref{fig:XRlc}). Conversely, the active X-ray softening of V404\,Cygni was much more rapid, occurring over $<$3d \citep{Plotkin-2017a}. Considering the strong relation between \lx\ and $\Gamma$, this softening timescale can be considered to be a proxy for the timescale of the luminosity decay.

Besides \maxiJ\ and V404\,Cygni, we identify one other source in the literature with comparable coverage of the X-ray softening and plateau of $\Gamma$: 4U\,1543$-$47, which softened from $\Gamma\approx1.6$ to $2.0$ over $\sim$5d and then plateaued during the decay of its 2002 outburst \citep{Kalemci-2005}. Another BH-LMXB, Swift\,J1357.2$-$0933, also has excellent spectral constraints of the X-ray softening during its 2011 and 2017 decays, which took $\sim$90 days \citep{Armas-Padilla-2013a,Beri-2019}, but became too faint to obtain useful constraints during the plateau stage. Considering the ensemble of these four systems, BH transitions into quiescence appear to fall into categories with gradual softenings (i.e., \maxiJ\ and Swift\,J1357.2$-$0933) and fast softenings (i.e., V404\,Cygni and 4U\,1543$-$47). One might be tempted to link the softening timescale to the $P_{\rm orb}$ of the system (and therefore the disk size for Roche Lobe overflow systems), since Swift\,J1357.2$-$0933 with the slowest softening ($\sim90$d) has the shortest orbital period \citep[$P_{\rm orb}=2.8$h;][]{Corral-Santana-2013} and V404\,Cygni with the fastest softening ($<$3d) has the longest orbital period \citep[$P_{\rm orb}=6.5$d;][]{Casares-1992a}, with \maxiJ\ \citep[$P_{\rm orb}=16.5$h;][]{Torres-2019b} and 4U\,1543$-$47 \citep[$P_{\rm orb}=1.1$d;][]{Orosz-1998} falling in between. Nevertheless, the apparent differences in the softening timescales between these four systems further highlights the need for improved X-ray spectral coverage during decays of systems spanning a wide range of properties like $P_{\rm orb}$, inclination, donor mass, etc., in order to attach a physical scenario to the timescale of the transition into quiescence.

\subsection{Radio Luminosity and Spectral Evolution}
\label{sec:radio_evo}

\begin{figure*}
    \centering
        \includegraphics[width=0.8\textwidth,trim={0 0 0 10mm}, clip]{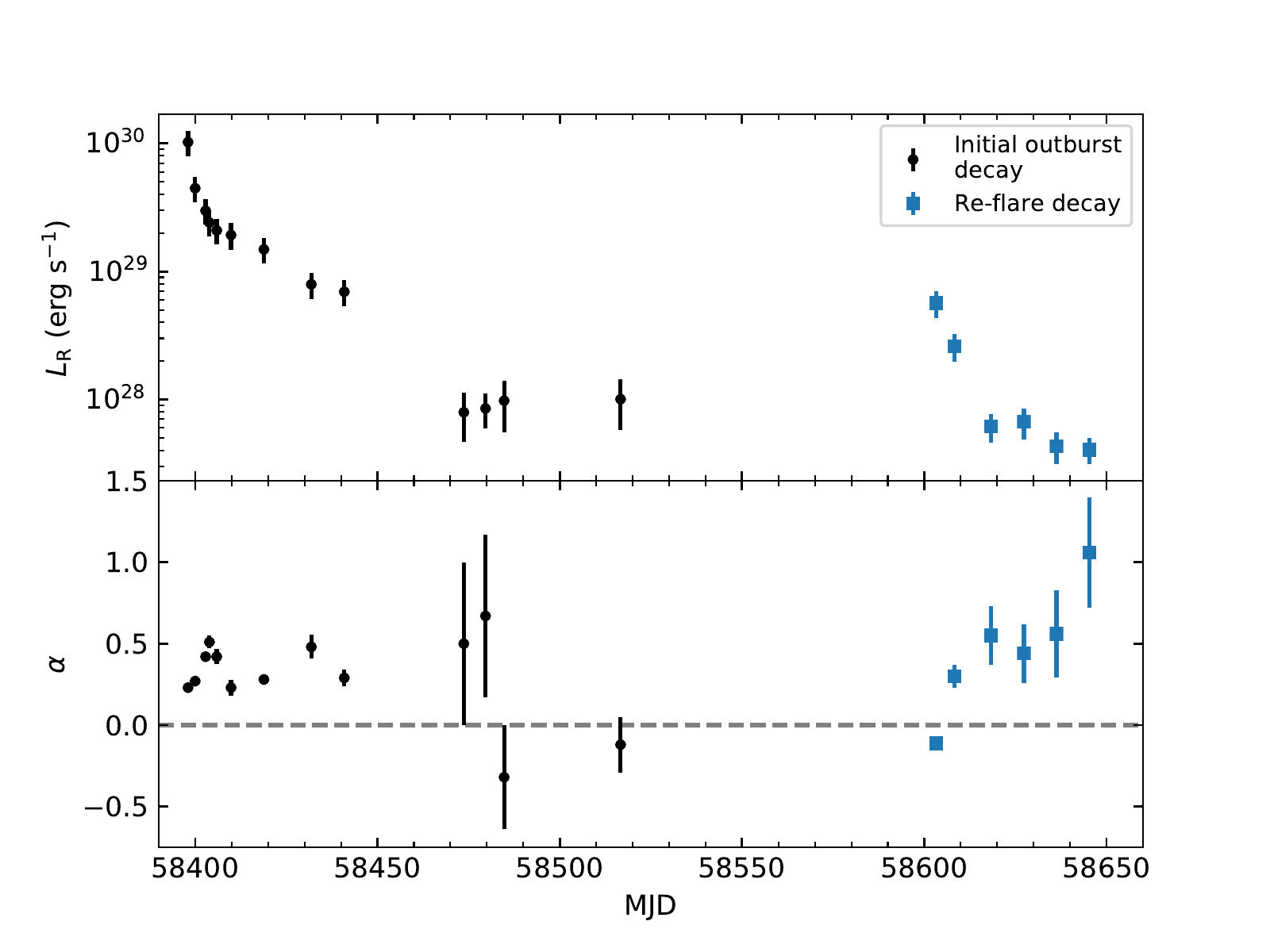}
        \caption{{\em Upper panel:} 5GHz radio light curve of \maxiJ. Luminosities were calculated by adopting the distance $d=2.96\pm0.33$ kpc as measured by \citet{Atri-2020}. {\em Lower panel:} The best-fit value of the radio spectral index ($\alpha$) for each observation. The horizontal dashed line represents a value of $\alpha=0$, i.e. a flat spectrum. In both panels, black, circular points represent observations of the decay of the initial outburst and blue, square points represent observations taken during the decay of the first re-flare.}
    \label{fig:Radlc}%
\end{figure*}

In the upper panel of Figure~\ref{fig:Radlc}, we show the 5 GHz radio light curve of the compact core of \maxiJ, and we show the time evolution of the radio spectral index $\alpha$ in the lower panel. The radio spectrum is inverted in nearly all of our observations (typically $\alpha > 0.2$; the weighted average spectral index is $\bar\alpha=0.24 \pm 0.06$, where the quoted uncertainty represents the standard deviation about the weighted average).

We see some variations in $\alpha$ with time. To investigate a potential correlation between $\log_{10}$\lr\ and $\alpha$, we calculate the Spearman correlation coefficient, $\rho$, and incorporate the uncertainties on the input parameters by simulating 10,000 datasets based on the original sample, allowing \lr\ and $\alpha$ to vary within their error bars according to a (log)normal distribution. We find $\rho=-0.2\pm0.2$, where this value represents the median of the 10,000 values of $\rho$ calculated from the simulated data, and the $1\sigma$ uncertainties are calculated as the 16th and 84th percentiles. The value of $\rho$ is consistent with no correlation between $\alpha$ and $\log_{10}$\lr. %, but could be indicative of a mild anti-correlation. 
To test the significance we adopt a Monte Carlo method by randomly shuffling the data, creating 10,000 new $\log_{10}$\lr, $\alpha$ pairs and calculating $\rho$ each time. We find that 17\% of the time we measure an anti-correlation that is as strong, or stronger than, $\rho=-0.2$ implying that our measured value of $\rho$ is not significant.

Perhaps the most interesting variation in $\alpha$ occurs on MJD 58603, where $\alpha$ veers slightly negative ($\alpha=-0.11 \pm 0.04$), and possibly also on MJDs 58485 and 58517 (at much lower statistical significance). Such negative deviations, however, are not exceptional and do not approach becoming optically thin (where typically $\alpha \approx -0.7$), and similarly, mild negative deviations in $\alpha$ have been observed for other low-luminosity hard state and quiescent BH-LMXBs \citep[e.g.][]{Rana-2016,Espinasse-2018,Plotkin-2019}. A slightly negative value of $\alpha$ such as the instances discussed here could perhaps indicate the jet break temporarily moving below $\sim$GHz radio frequencies.  Alternatively, we may be seeing deviations from a simple conical jet, or we may be seeing an optically thin emission region (perhaps related to a small flare) superposed over a flat/inverted radio spectrum from the compact jet.% and may be related to the jet recovering from a small flare \citep[see e.g.][]{Rana-2016,TetarenkoA-2019,Plotkin-2019}.

\subsection{Radio --- X-ray Correlation}

\begin{figure*}
    \centering
    \includegraphics[width=0.9\textwidth]{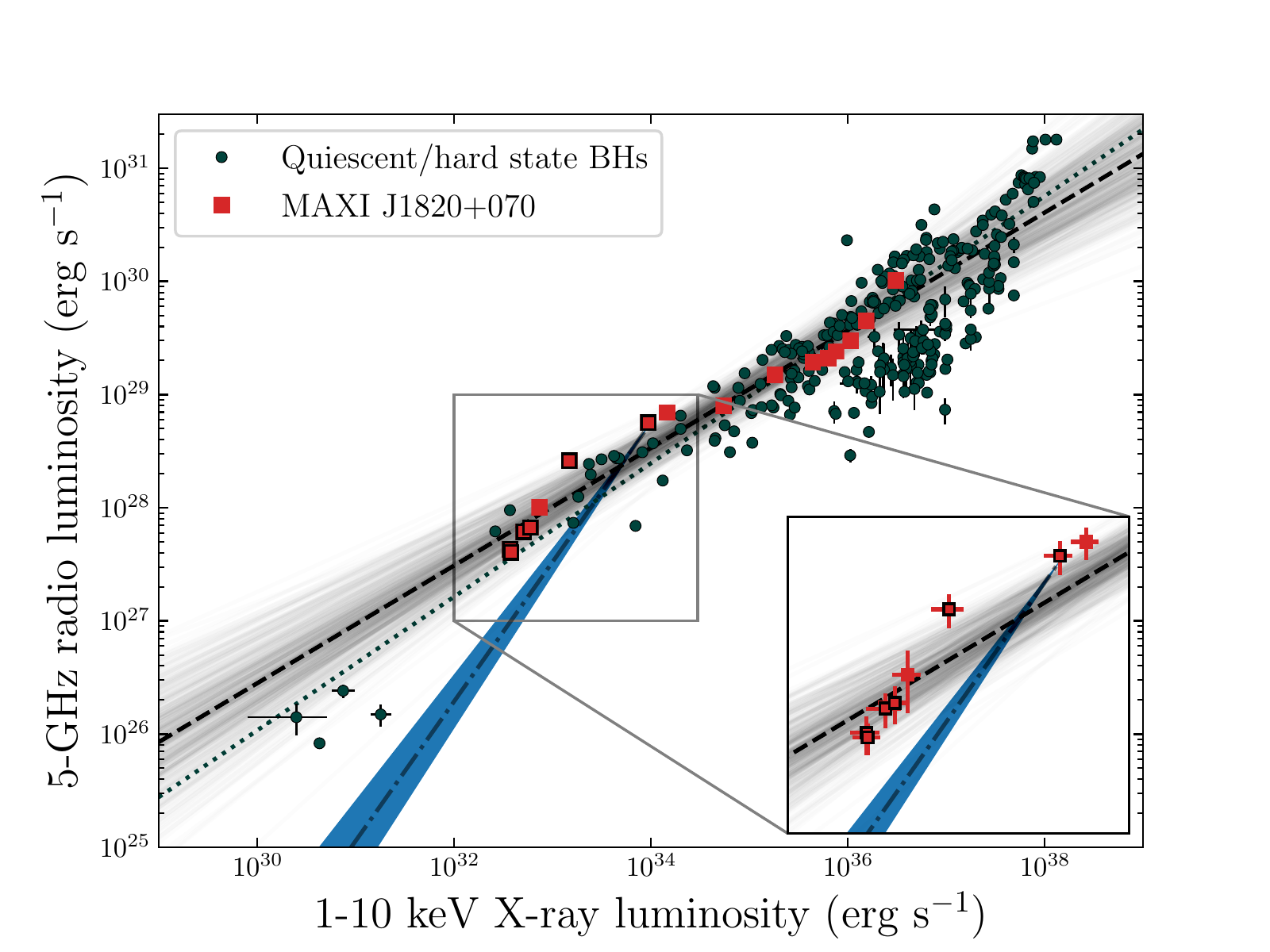}
    \caption{\maxiJ\ on the radio --- X-ray plane (red squares). Observations taken during the decay of the first re-flare are highlighted with black boxes. Also plotted for comparison are a sample of quiescent/hard-state BH-LMXBs (dark green circles). The dashed line shows the best-fit \lr\ $\propto$ \lx$^{0.52}$ correlation to the \maxiJ\ data. For clarity, in the main panel we do not plot the uncertainties on the \maxiJ\ luminosities, but they are taken into account during the fitting process. The grey solid lines show some samples from the posterior distribution of best fits to the correlation. The dotted line shows the best-fit \lr\ $\propto$ \lx$^{0.59}$ correlation for a sample of 36 BHs from \citet{Gallo-2018}, highlighting the `standard track.' The dot-dashed line indicates the predicted path that \maxiJ\ would take through the radio --- X-ray plane if the observed spectral softening (which completes at \lx $\approx 10^{34}$ erg s$^{-1}$) was due to a switch to synchrotron-cooled X-ray emission, with the blue shaded region representing the expected scatter about the line. We show a zoomed in region of the lowest luminosity points inset, with error bars on the data. Data for the quiescent/hard-state BHs were obtained from A. Bahramian's Radio/X-ray correlation database for X-ray binaries \citep{Bahramian-2018-LrLx}.}
    \label{fig:LrLx}
\end{figure*}

%LrLx table goes here

\begin{deluxetable*}{cccccc}
\label{tab:lrlx}
\tablecaption{Summary of results from the quasi-simultaneous X-ray and radio observations}
\tablewidth{0pt}
\tablehead{
\colhead{MJD$_{\rm X}$} &
\colhead{$\log_{10}L_{\rm X}$(1--10 keV)\tablenotemark{a}} &
\colhead{$\Gamma$} &
\colhead{MJD$_{\rm R}$} &
\colhead{$\log_{10}L_{\rm R}$(5 GHz)} &
\colhead{$\alpha$}
\\
\colhead{} &
\colhead{(erg s$^{-1}$)} &
\colhead{} &
\colhead{} &
\colhead{(erg s$^{-1}$)} &
\colhead{}
}
\decimalcolnumbers

\startdata
$58397.09$ & $36.5\pm0.1$ & $1.66\pm0.02$ & $58398.04$ & $30.0\pm0.1$ & $0.23\pm0.00$ \\ 
$58400.01$ & $36.2\pm0.1$ & $1.63\pm0.03$ & $58399.99$ & $29.6\pm0.1$ & $0.27\pm0.01$ \\ 
$58402.28$ & $36.0\pm0.1$ & $1.59\pm0.04$ & $58402.85$ & $29.5\pm0.1$ & $0.42\pm0.02$ \\ 
$58404.26$ & $35.9\pm0.1$ & $1.63\pm0.02$ & $58403.91$ & $29.4\pm0.1$ & $0.51\pm0.04$ \\ 
$58406.73$ & $35.8\pm0.1$ & $1.60\pm0.02$ & $58405.90$ & $29.3\pm0.1$ & $0.42\pm0.05$ \\ 
$58410.30$ & $35.6\pm0.1$ & $1.65\pm0.02$ & $58409.83$ & $29.3\pm0.1$ & $0.23\pm0.05$ \\ 
$58419.61$ & $35.3\pm0.1$ & $1.67\pm0.01$ & $58418.85$ & $29.2\pm0.1$ & $0.28\pm0.03$ \\ 
$58432.55$ & $34.7\pm0.1$ & $1.74\pm0.14$ & $58431.85$ & $28.9\pm0.1$ & $0.48\pm0.07$ \\ 
$58441.11$ & $34.2\pm0.1$ & $1.84\pm0.07$ & $58440.90$ & $28.8\pm0.1$ & $0.29\pm0.05$ \\ 
$58518.61$ & $32.9\pm0.1$ & $1.84\pm0.10$ & $58516.70$ & $28.0\pm0.2$ & $-0.12\pm0.17$ \\ 
$58603.13$ & $34.0\pm0.1$ & $1.97\pm0.14$ & $58603.32$ & $28.8\pm0.1$ & $-0.11\pm0.04$ \\ 
$58609.33$ & $33.2\pm0.1$ & $2.19\pm0.23$ & $58608.35$ & $28.4\pm0.1$ & $0.30\pm0.07$ \\ 
$58618.67$ & $32.7\pm0.1$ & $1.99^{+0.24}_{-0.23}$ & $58618.34$ & $27.8\pm0.1$ & $0.55\pm0.18$ \\ 
$58627.42$ & $32.8\pm0.1$ & $1.95\pm0.15$ & $58627.34$ & $27.8\pm0.1$ & $0.44\pm0.18$ \\ 
$58636.67$ & $32.6\pm0.1$ & $2.11\pm0.12$ & $58636.26$ & $27.6\pm0.1$ & $0.56\pm0.27$ \\ 
$58645.08$ & $32.6\pm0.1$ & $2.04\pm0.07$ & $58645.24$ & $27.6\pm0.1$ & $1.06\pm0.34$ \\  
\enddata
\tablenotetext{a}{Unabsorbed}
\end{deluxetable*}

Figure~\ref{fig:LrLx} shows \maxiJ\ in the \lr\ --- \lx\ plane. %, along with a sample of BHs compiled by \citet{Bahramian-2018-LrLx}. 
\lr\ and \lx, along with their corresponding spectral indices, are also tabulated in Table \ref{tab:lrlx}. \maxiJ\ displays a non-linear correlation over $\approx$4 decades in X-ray luminosity extending from the hard state through quiescence. To measure the slope of the correlation we adopt the Bayesian modeling package {\tt linmix}\footnote{\href{https://linmix.readthedocs.io/en/latest/}{https://linmix.readthedocs.io/en/latest/}}, the {\sc python} port of the {\sc linmix\_err} {\sc idl} package \citep{Kelly-2007} to measure the dependence of \lr\ on \lx. We assume a linear correlation of the form $\ell_{\rm R} = b + m\ell_{\rm X}$, where $\ell_{\rm R}$, $\ell_{\rm X}$ are the logarithms of \lr\ and \lx, respectively and $b$ and $m$ are the $\ell_{\rm R}$-intercept and slope, respectively. Following \citet{Gallo-2012,Gallo-2014b,Gallo-2018}, we include additional uncertainties of 0.3 dex on $\ell_{\rm R}$ and $\ell_{\rm X}$ to account for lack of strict simultaneity between our radio and X-ray observations. \maxiJ\ follows a relation with a slope $m=0.52\pm0.07$, with an intrinsic random scatter $\sigma_0=0.14^{+0.10}_{-0.07}$ dex, where the best-fit values represent the median of 10,000 draws from the posterior distributions and the $1\sigma$ uncertainties are calculated as the 16th and 84th percentiles. The correlation slope is consistent with that derived for the larger hard state BH population \citep[$m=0.59\pm0.02$ from 36 BHs;][]{Gallo-2018}.

\maxiJ\ is only the fourth individual source for which an unbroken \lr\ --- \lx\ correlation has been tracked over four (or more) decades in X-ray luminosity, joining V404\,Cygni ($m=0.54\pm0.03$; \citealt{Corbel-2008, Plotkin-2017a}), GX\,339$-$4 ($m=0.62\pm0.01$; \citealt{Corbel-2013a}; see also \citealt{Tremou-2020}), and XTE\,J1118$+$480 ($m=0.72\pm0.09$; \citealt{Gallo-2014b}). Note, we are aware of two other sources with \lr\ --- \lx\ coverage over a comparable range of luminosities, H1743$-$322 and Swift J1753.5$-$0127;  however, both sources appear to follow different correlation slopes above and below \lx$\approx 10^{34}$ erg s$^{-1}$ \citep{Coriat-2011, Plotkin-2017b}, such that we do not compare \maxiJ\ to those two sources here. \maxiJ\ also shows the same \lr\ --- \lx\ correlation over multiple decays, similar to V404\,Cygni \citep{Plotkin-2017a} and GX\,339$-$4 \citep{Corbel-2013a}. This suggests that the coupling between the disk and the jet is robust between outburst decays.

We can combine the slope of the \lr\ --- \lx\ correlation with radio spectral information to draw some conclusions about the dominant emission mechanisms at work as the source decays. If the radio luminosity is responding to changes in the mass accretion rate ($\dot{M}$), then \lr $\propto \dot{M}^{\frac{17}{12}-\frac{2}{3}\alpha}$ \citep{Heinz-2003, Markoff-2003}.\footnote{\citet{Heinz-2003} define $f_\nu \propto \nu^{-\alpha}$, so the formulae in this work differ by a minus sign.} The above assumes that jets are emitting synchrotron radiation from non-thermal particles accelerated into a power-law distribution, $dn/d\gamma \propto \gamma^{-p}$ (where $n$ is the number density of particles, $\gamma$ is the Lorentz factor, and $p$ is the power-law index that describes the energy distribution of the particles.  Following \citealt{Heinz-2003}, we adopt $p=2$). %For the \citet{Heinz-2003} prescription to apply, we assume $p=2$.} 
For an arbitrary X-ray emission process, we can assume \lx$\propto \dot{M}^q$, where $q$ is a number that parameterizes the radiative efficiency of the X-ray process\footnote{We make the assumption that the bolometric correction is constant with changing $\dot{M}$} (i.e., $q \neq 1$ indicates a non-thermal or inefficient mechanism). Thus, the measured slope of the \lr\ --- \lx\ correlation can be written as

\begin{equation}
\label{eq:beta}
    m = \frac{\frac{17}{12} - \frac{2}{3}\alpha}{q}.
\end{equation}

\noindent For $m=0.52\pm0.07$ and $\bar\alpha=0.24\pm0.06$ (i.e., the weighted average radio spectral index during our campaign), we find, on average during the two decays, that $q=2.4\pm0.3$. Thus, we require a radiatively inefficient source of X-ray emission. 

Given our empirical constraint on the radiative efficiency (through the $q$ parameter), we can make some inferences on the most likely source of X-ray emission. Inverse Comptonization and/or SSC processes in a RIAF can readily produce the observed X-ray signatures.  As the accretion rate drops, one expects the density of the inner region of the accretion flow to decrease.  In turn, the optical depth to inverse Compton scattering may then decrease and yield a gradual increase in $\Gamma$ \citep[e.g.,][]{Esin-1997}.

In terms of jet origins for X-ray emission, we describe below that most scenarios are unlikely, except perhaps in some instances of jet-related SSC emission, which is difficult to distinguish from SSC originating from a RIAF. As expanded upon below, different mechanisms for (jet) X-ray emission are expected to emit with different radiative efficiencies.  Thus, in the following we walk through different scenarios for jet emission, and in order for a scenario to be deemed viable its predictions must be consistent with our observations on the average radio spectral index, with the observed evolution in $\Gamma$, \textit{and} with the behavior in \lr\ --- \lx.

Since we have a precise measurement of the average radio spectral index, our combined constraints on the X-ray spectral softening and \lr\ --- \lx\ allow us to exclude other scenarios for jet-dominated X-rays. Although, the following discussion comes with the caveat that we assume that changes in radio/X-ray luminosities are driven primarily by changes in $\dot{M}$. If we assume that X-rays are always dominated by optically thin synchrotron radiation (again, emitted by a non-thermal electron population described by $dn/d\gamma \propto \gamma^{-p}$), then the X-ray spectral softening from $\Gamma=1.5$ to $2$ (as seen during the re-flare) could be explained as optically thin synchrotron emission emanating from particles evolving to a steeper non-thermal distribution ($p=2\Gamma-1$ evolving from 2 to 3). Then, according to Equation 17c of \citet{Heinz-2003}, we would expect the slope of the \lr --- \lx\ correlation to evolve from 0.7 to 0.6 in coordination with the X-ray spectrum softening from $\Gamma = 1.5$ to 2.0 (again, adopting a radio spectral index of $\bar\alpha=0.24\pm0.06$).  Therefore, if $p$ were truly evolving to a steeper distribution, then we would expect to see the \lr --- \lx\ slope become shallower as the source decays to quiescence, an effect that does not appear to be seen in Fig. \ref{fig:LrLx}.

From \lr\ --- \lx\ we can also exclude the possibility that the X-ray spectral softening is caused by a synchrotron cooled jet dominating the X-rays. If this were the case we would expect the transition to occur at \lx$\approx10^{34}$ erg s$^{-1}$ (i.e. the luminosity at which $\Gamma$ reaches 2), at which point the radiative efficiency would be $q_{\rm cool} = p + 2 - \left(\frac{3}{2}\right)\Gamma$ (\citealt{Heinz-2004}, see also section 4.2 of \citealt{Plotkin-2017a}). For $p=2$ as expected for a distribution of synchrotron emitting particles for which we measure $\Gamma=1.5$ at its hardest \citep[$p=2\Gamma-1$; see also e.g.][]{Heinz-2003}, the radiative efficiency would be $q_{\rm cool}=1$. Adopting the weighted average value of $\bar\alpha=0.24\pm0.06$ and the weighted average of all measured values of $\Gamma$ at luminosities \lx$\leq10^{34}$ erg s$^{-1}$ ($\Gamma=1.99\pm0.09$), we would therefore expect to see the slope of the \lr\ --- \lx\ correlation increase to $m=1.23\pm0.12$ at \lx$=10^{34}$ erg s$^{-1}$ \citep{Yuan-2005}. In Figure~\ref{fig:LrLx} we show that all of the lowest radio luminosity data points lie above the expected synchrotron cooling decay, so we can likely rule it out.

Though we have ruled out the majority of pure jet models for the X-ray emission, we still identify a jet origin for SSC emission as a possibility. Jet SSC emission could plausibly come from either a thermal or a non-thermal particle distribution. A softening might then be expected if the Comptonized spectrum is produced by single scatterings of synchrotron-cooled seed photons \citep[see e.g.][for discussions]{Corbel-2008,Plotkin-2017a}. However, confirming or refuting this scenario requires numerical modeling that is out of the scope of this paper. This would allow us to understand how the extra source of cooling would affect particle energies and therefore the power and spectrum of the seed synchrotron photons, as well as the average number of scatterings before inverse Comptonized photons escape, etc. In lieu of such modeling, for the time being we believe a RIAF origin for the X-ray emission is the most reasonable.

\subsubsection{Deviations from \lr\ --- \lx}

\begin{figure}
    \centering
    \includegraphics[width=0.48\textwidth, trim=5mm 0 5mm 2mm, clip]{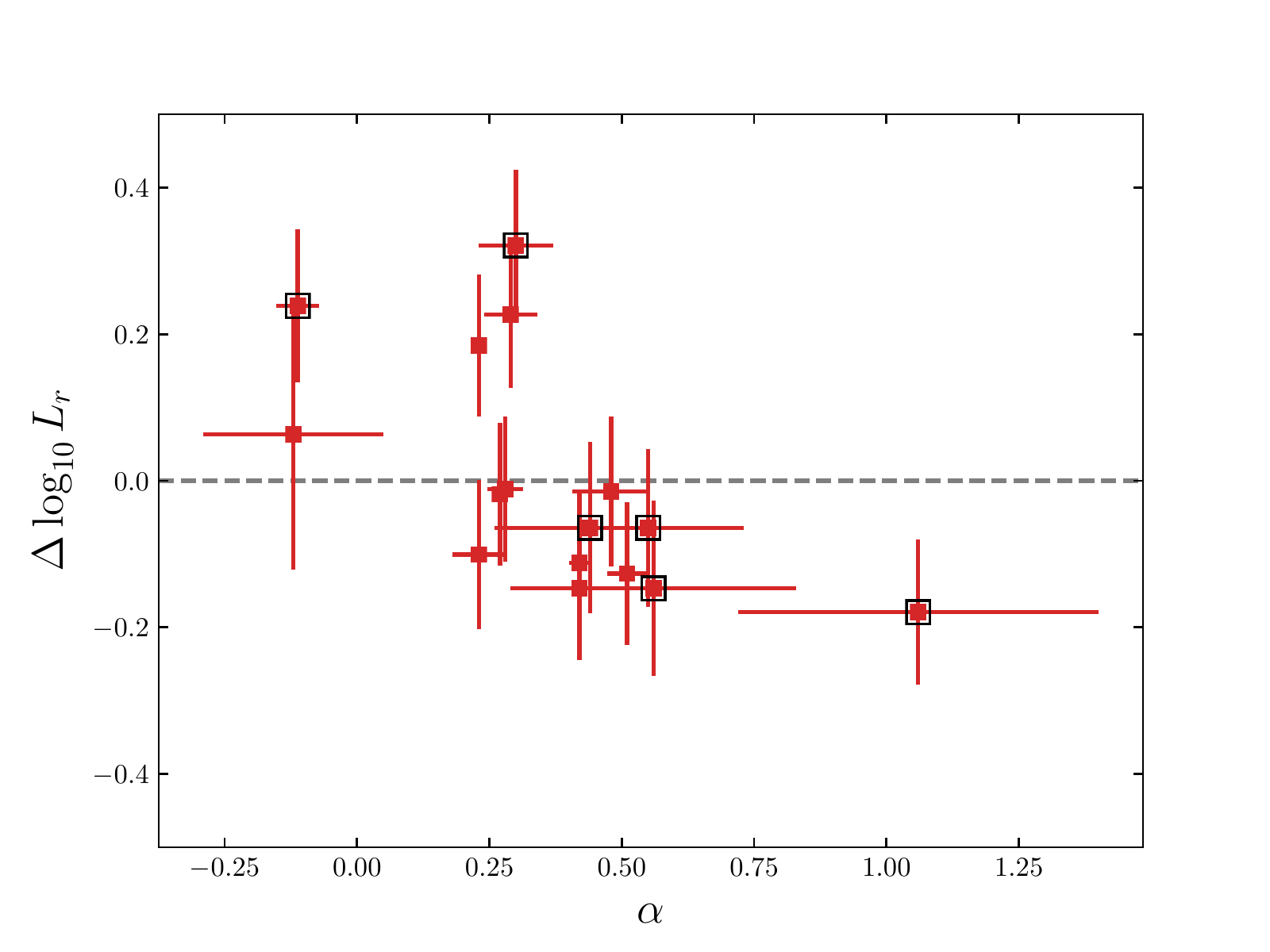}
    \caption{Logarithmic radio luminosity residuals vs. radio spectral index. We define the residuals as the difference between $\log_{10}$\lr\ and its expected (logarithmic) value according to the best-fit \lr\ --- \lx\ correlation. %We calculate the residuals on a logarithmic scale. 
    As in Figure~\ref{fig:LrLx}, the black boxes highlight the observations taken during the decay of the first re-flare.}
    \label{fig:residuals}
\end{figure}

We stress that our best-fit slope only describes the \textit{average} path taken through the \lr\ --- \lx\ plane, since we observe temporary deviations about our best-fit line (which our regression technique models as intrinsic scatter). %The variance of the residuals in \lr\ (where we define the residuals as data/model) is $\sigma^2=0.17$. 
Similar deviations have also been seen for GX 339--4, and they illustrate the importance of measuring \lr\ --- \lx\ slopes over a wide dynamic range in luminosity (see Section 4.2.3 of \citealt{Corbel-2013a} for a detailed discussion). In Figure~\ref{fig:residuals} we plot the (logarithmic) radio luminosity residuals against $\alpha$ and find a marginal possibility of an anti-correlation. We calculate a Spearman correlation coefficient $\rho=-0.5\pm0.2$, which we find has a $\sim2$\% probability of occurring by chance in randomly shuffled data (see Section~\ref{sec:radio_evo} for our methodology).  

Although we do not consider this to be a statistically significant anti-correlation, it is an intriguing result that warrants further observational scrutiny (on future outburst decays). Deviations from a simple \citet{Blandford-1979} compact jet can result in changes in $\alpha$, which would then cause an anti-correlation between $\alpha$ and $m$  (e.g., according to Equation~\ref{eq:beta}, if $\alpha$ increases then $m$ decreases; also see \citealt{Corbel-2013a} for a discussion on how $\alpha$ can influence a source's position in the \lr --- \lx\ plane). Thus, if the anti-correlation is indeed real, it may not be unexpected,  and it could be manifesting itself as the (observed) deviations from the global \lr\ --- \lx\ correlation.  Unfortunately we cannot quantify if $\alpha$ really is the driver because we are not able to make precise enough measurements of the slopes of individual deviations (since they cover too little dynamic range in luminosity) to directly compare changes in $m$  to changes in $\alpha$.  Nevertheless, Figure~\ref{fig:residuals} motivates the importance of obtaining meaningful constraints on $\alpha$ during high-cadence multiwavelength monitoring of future outburst decays.

\section{Summary and Conclusions}
\label{sec:conclusions}

We presented high-cadence, quasi-simultaneous X-ray and radio spectral monitoring of the BH-LMXB \maxiJ\ during the decline of its 2018 outburst and of a subsequent re-flare. We find that, similar to previous observations of BH-LMXBs as they approach quiescence, the X-ray spectrum softened from $\Gamma\sim1.7$ to $\Gamma\sim2$ over the course of $\sim30$d, `saturating' at $\Gamma\sim2$ even as the luminosity continued to decrease below \lx$\lesssim10^{34}$ erg s$^{-1}$ ($10^{-5}$ \ledd). During this time, the radio spectrum generally remained slightly inverted, as the source luminosity decreased, with occasional deviations to a negative slope. However, the radio spectrum never reached values indicative of optically thin synchrotron emission.

We find that \maxiJ\ follows a correlation in the \lr\ --- \lx\ plane of the form \lr$\propto$\lx$^{0.52\pm0.07}$, meaning that it is the fourth BH-LMXB, after GX\,339$-$4, XTE\,J1118$+$480 and V404\,Cygni, to follow the standard track over $\gtrsim4$ decades in \lx. We use the slope of the \lr\ --- \lx\ correlation, along with the average radio spectral index $\bar\alpha=0.24\pm0.06$, to estimate the radiative efficiency parameter $q$. We find, on average, $q=2.4\pm0.3$ over the two decays, suggesting that the X-ray emission is produced by a radiatively inefficient mechanism. The calculated value of $q$ allows us to rule out particle acceleration along the jet becoming less efficient as the emission process responsible for the observed X-ray softening. The fact that the \lr\ --- \lx\ slope remains steady and does not steepen at $10^{34}$ erg s$^{-1}$ also allows us to effectively exclude a synchrotron-cooled jet as the cause of the softening. 

Excluding these jet origins for the X-ray emission implies that the dominant mechanism responsible for the observed softening is either Comptonization processes happening within a RIAF, or possibly SSC processes from within the jet. \maxiJ\ is one of only two BH-LMXBs (the other being V404\,Cygni) that have both good coverage in the \lr\ --- \lx\ plane {\em and} a well constrained distance \citep{Atri-2020}, making this an important data set for studying BH-LMXBs during their decay to quiescence, as we have strong constraints on its luminosity. In the future, high spectral quality, strictly simultaneous radio and X-ray observations of other {\em nearby} transients (with low column density) are needed to eventually build population studies, in order to understand which details of the transition to quiescence are universal, and which details instead depend on system parameters (e.g., orbital period, donor mass, etc.) and disk/jet couplings (i.e., radio/X-ray luminosity correlation slope) that are specific to individual systems.

\acknowledgments

We thank the anonymous referee for useful comments that have helped improve the manuscript. AWS would like to thank Bailey Tetarenko for useful discussions regarding the X-ray decay. AWS would also like to thank Arash Bahramian for providing access to their exhaustive library of \lr\ --- \lx\ data for BH-LMXBs. JCAM-J was the recipient of an Australian Research Council Future Fellowship (FT140101082), funded by the Australian government. PK, EG and JH acknowledge financial support that was provided by the National Aeronautics and Space Administration through \cha\ Award Numbers GO8-19033X, GO8-19027B and GO9-20027B issued by the \cha\ X-ray Center, which is operated by the Smithsonian Astrophysical Observatory for and on behalf of the National Aeronautics Space Administration under contract NAS8-03060. This work made use of data supplied by the UK Swift Science Data Centre at the University of Leicester. The scientific results reported in this article are based to a significant degree on observations made by the Chandra X-ray Observatory. This research has made use of data and software provided by the High Energy Astrophysics Science Archive Research Center (HEASARC), which is a service of the Astrophysics Science Division at NASA/GSFC. The National Radio Astronomy Observatory is a facility of the National Science Foundation operated under cooperative agreement by Associated Universities, Inc.

\facilities{CXO, NICER, Swift, VLA}

\software{CASA \citep{McMullin-2007}, CIAO \citep{Fruscione-2006}, HEASoft (\href{https://heasarc.gsfc.nasa.gov/lheasoft/}{https://heasarc.gsfc.nasa.gov/lheasoft/}), LinMix \citep[][\href{https://linmix.readthedocs.io/en/latest/}{https://linmix.readthedocs.io/en/latest/}]{Kelly-2007}, UVMULTIFIT \citep{Marti-Vidal-2014}, XSPEC \citep{Arnaud-1996}}

\bibliography{references}{}
\bibliographystyle{aasjournal}

\appendix 
\restartappendixnumbering
\section{Observation summary tables}
\label{sec:obs_tables}

\setlength{\tabcolsep}{3pt}
\startlongtable
\begin{deluxetable*}{llcccccc}
\label{tab:X-ray_obs}
\tablecaption{Summary of X-ray observations and best-fit spectral measurements}
\tablewidth{0pt}
\tablehead{
\colhead{Instrument} &
\colhead{ObsID} &
\colhead{MJD\tablenotemark{a}} &
\colhead{$t_{\rm exp}$} &
\colhead{Net Count Rate\tablenotemark{b}} &
\colhead{$\Gamma$} &
\colhead{$\log_{10}L_{\rm X}$(1--10 keV)} &
\colhead{$W$/dof\tablenotemark{c}}
\\
\colhead{} &
\colhead{} &
\colhead{} &
\colhead{(ks)} &
\colhead{(count s$^{-1}$)} &
\colhead{} &
\colhead{(erg s$^{-1}$)} &
\colhead{}
}
\decimalcolnumbers

\startdata
Swift/XRT-WT & 00010627109 & $58397.09$ & $1.0$ & $80.52$ & $1.66\pm0.02$\tablenotemark{d} &$36.5\pm0.1$ & $654/640$ \\ 
Swift/XRT-WT & 00010627110 & $58400.01$ & $0.6$ & $37.09$ & $1.63\pm0.03$\tablenotemark{e} &$36.2\pm0.1$ & $470/461$ \\ 
Swift/XRT-WT & 00010627111 & $58402.28$ & $0.6$ & $24.41$ & $1.59\pm0.04$\tablenotemark{f} &$36.0\pm0.1$ & $401/383$ \\ 
Swift/XRT-WT & 00010627112 & $58404.26$ & $1.0$ & $12.98$ & $1.63\pm0.02$ &$35.9\pm0.1$ & $397/375$ \\ 
Swift/XRT-WT & 00010627113 & $58406.73$ & $0.9$ & $13.85$ & $1.60\pm0.02$ &$35.8\pm0.1$ & $433/375$ \\ 
Swift/XRT-WT & 00010627114 & $58408.25$ & $1.0$ & $12.64$ & $1.58\pm0.02$ &$35.8\pm0.1$ & $360/372$ \\ 
Swift/XRT-WT & 00010627115 & $58410.30$ & $1.3$ & $10.21$ & $1.65\pm0.02$ &$35.6\pm0.1$ & $426/381$ \\ 
Swift/XRT-WT & 00010627116 & $58412.03$ & $1.7$ & $7.94$ & $1.58\pm0.02$ &$35.6\pm0.1$ & $392/390$ \\ 
Swift/XRT-WT & 00010627119 & $58417.48$ & $6.9$ & $4.03$ & $1.64\pm0.02$ &$35.4\pm0.1$ & $470/503$ \\ 
Swift/XRT-WT & 00010627120 & $58419.61$ & $12.5$ & $3.91$ & $1.67\pm0.01$ &$35.3\pm0.1$ & $634/578$ \\ 
Swift/XRT-WT & 00010627121 & $58422.27$ & $2.0$ & $3.41$ & $1.70\pm0.03$ &$35.2\pm0.1$ & $296/271$ \\ 
Swift/XRT-WT & 00010627122 & $58424.66$ & $2.0$ & $2.61$ & $1.73\pm0.04$ &$35.1\pm0.1$ & $264/234$ \\ 
Swift/XRT-WT & 00010627123 & $58425.46$ & $1.4$ & $2.58$ & $1.74\pm0.05$ &$35.0\pm0.1$ & $163/183$ \\ 
Swift/XRT-WT & 00010627124 & $58426.65$ & $0.7$ & $2.49$ & $1.74\pm0.07$ &$35.0\pm0.1$ & $75/100$ \\ 
Swift/XRT-WT & 00010627125 & $58428.17$ & $2.2$ & $2.12$ & $1.77\pm0.04$ &$34.9\pm0.1$ & $202/222$ \\ 
Swift/XRT-WT & 00010627126 & $58430.70$ & $1.6$ & $0.90$ & $1.79\pm0.08$ &$34.8\pm0.1$ & $92/83$ \\ 
Swift/XRT-WT & 00010627127 & $58432.55$ & $0.4$ & $1.27$ & $1.74\pm0.14$ &$34.7\pm0.1$ & $29/31$ \\ 
Swift/XRT-WT & 00010627128 & $58434.35$ & $2.0$ & $1.09$ & $1.80\pm0.06$ &$34.6\pm0.1$ & $146/119$ \\ 
Chandra/ACIS-S & 20207 & $58435.77$ & $0.3$\tablenotemark{g} & $3.08$ & $1.73\pm0.10$ &$34.6\pm0.1$ & $66/52$ \\ 
Swift/XRT-WT & 00010627129 & $58436.53$ & $1.9$ & $0.74$ & $1.75\pm0.08$ &$34.5\pm0.1$ & $95/82$ \\ 
Swift/XRT-WT & 00010627130 & $58438.59$ & $1.6$ & $0.57$ & $1.79\pm0.10$ &$34.4\pm0.1$ & $44/59$ \\ 
Swift/XRT-WT & 00010627131 & $58440.19$ & $1.7$ & $0.53$ & $1.70^{+0.11}_{-0.10}$ &$34.3\pm0.1$ & $60/55$ \\ 
Chandra/ACIS-S & 20186 & $58441.11$ & $8.2$ & $0.19$ & $1.84\pm0.07$ &$34.2\pm0.1$ & $416/438$ \\ 
NICER/XTI & 1200120310 & $58441.26$ & $0.5$ & $8.42$ & $1.88\pm0.04$ &$34.3\pm0.1$ & $148/128$ \\ 
NICER/XTI & 1200120312 & $58443.84$ & $1.2$ & $5.22$ & $1.97\pm0.04$ &$34.0\pm0.1$ & $198/176$ \\ 
Chandra/ACIS-S & 20208 & $58518.61$ & $19.1$ & $0.06$ & $1.84\pm0.10$ &$32.9\pm0.1$ & $48/59$ \\ 
Chandra/ACIS-S & 22080 & $58519.12$ & $19.1$ & $0.04$ & $1.96\pm0.11$ &$32.7\pm0.1$ & $37/47$ \\ 
Swift/XRT-WT & 00010627145 & $58563.12$ & $1.1$ & $19.46$ & $1.54\pm0.02$ &$36.0\pm0.1$ & $523/472$ \\ 
Swift/XRT-WT & 00010627146 & $58564.25$ & $1.0$ & $16.78$ & $1.49\pm0.02$ &$35.9\pm0.1$ & $421/444$ \\ 
Swift/XRT-WT & 00010627147 & $58565.18$ & $1.0$ & $14.13$ & $1.53\pm0.02$ &$35.9\pm0.1$ & $422/408$ \\ 
Swift/XRT-WT & 00010627148 & $58566.24$ & $2.0$ & $16.32$ & $1.51\pm0.01$ &$35.9\pm0.1$ & $550/555$ \\ 
Swift/XRT-WT & 00010627149 & $58567.17$ & $2.1$ & $14.51$ & $1.55\pm0.02$ &$35.9\pm0.1$ & $525/531$ \\ 
Swift/XRT-WT & 00010627150 & $58571.03$ & $0.6$ & $12.29$ & $1.52\pm0.03$ &$35.8\pm0.1$ & $276/298$ \\ 
Swift/XRT-WT & 00010627151 & $58572.15$ & $1.0$ & $11.83$ & $1.58\pm0.02$ &$35.7\pm0.1$ & $398/365$ \\ 
Swift/XRT-WT & 00010627152 & $58573.34$ & $0.7$ & $11.28$ & $1.62\pm0.03$ &$35.7\pm0.1$ & $286/301$ \\ 
Swift/XRT-WT & 00010627153 & $58574.33$ & $1.0$ & $11.09$ & $1.62\pm0.03$ &$35.7\pm0.1$ & $402/357$ \\ 
Swift/XRT-WT & 00010627154 & $58575.14$ & $1.0$ & $10.21$ & $1.61\pm0.03$ &$35.7\pm0.1$ & $341/336$ \\ 
Swift/XRT-WT & 00010627155 & $58576.06$ & $0.1$ & $8.67$ & $1.64\pm0.08$ &$35.6\pm0.1$ & $73/73$ \\ 
Swift/XRT-WT & 00010627156 & $58577.05$ & $0.6$ & $8.49$ & $1.62\pm0.04$ &$35.6\pm0.1$ & $222/225$ \\ 
Swift/XRT-WT & 00010627158 & $58583.17$ & $0.7$ & $3.00$ & $1.61\pm0.06$ &$35.3\pm0.1$ & $140/118$ \\ 
Swift/XRT-WT & 00010627159 & $58584.70$ & $0.8$ & $4.08$ & $1.69\pm0.05$ &$35.2\pm0.1$ & $159/169$ \\ 
Swift/XRT-WT & 00010627161 & $58586.68$ & $0.9$ & $1.07$ & $1.66\pm0.10$ &$35.0\pm0.1$ & $45/58$ \\ 
Swift/XRT-WT & 00010627162 & $58588.81$ & $0.8$ & $2.54$ & $1.72\pm0.06$ &$35.0\pm0.1$ & $145/117$ \\ 
Swift/XRT-WT & 00010627163 & $58590.81$ & $1.0$ & $1.96$ & $1.63\pm0.06$ &$34.9\pm0.1$ & $130/113$ \\ 
Chandra/ACIS-S & 21200 & $58603.13$ & $3.7$ & $0.12$ & $1.97\pm0.14$ &$34.0\pm0.1$ & $207/260$ \\ 
Chandra/ACIS-S & 21201 & $58609.33$ & $8.2$ & $0.02$ & $2.19\pm0.23$ &$33.2\pm0.1$ & $127/126$ \\ 
Chandra/ACIS-S & 21202 & $58618.67$ & $4.8$ & $0.04$ & $1.99^{+0.24}_{-0.23}$ &$32.7\pm0.1$ & $103/122$ \\ 
Chandra/ACIS-S & 21203 & $58627.42$ & $11.4$ & $0.05$ & $1.95\pm0.15$ &$32.8\pm0.1$ & $25/31$ \\ 
Chandra/ACIS-S & 21204 & $58636.67$ & $28.2$ & $0.03$ & $2.11\pm0.12$ &$32.6\pm0.1$ & $44/50$ \\ 
Chandra/ACIS-S & 21205 & $58645.08$ & $65.6$ & $0.03$ & $2.04\pm0.07$ &$32.6\pm0.1$ & $121/103$ \\  
\enddata
\tablenotetext{a}{MJD at the start of the observation}
\tablenotetext{b}{The typical uncertainty on Net count rate is $\sim0.1$ count s$^{-1}$}
\tablenotetext{c}{\citet{Cash-1979} statistic, modified for background-subtracted spectra}
\tablenotetext{d}{Best-fit spectrum included a {\tt diskbb} component with $kT=0.18$ keV}
\tablenotetext{e}{Best-fit spectrum included a {\tt diskbb} component with $kT=0.13$ keV}
\tablenotetext{f}{Best-fit spectrum included a {\tt diskbb} component with $kT=0.17$ keV}
\tablenotetext{g}{Effective exposure time for the read-out streak}
\end{deluxetable*}

\begin{deluxetable*}{ccccccc}
\label{tab:radio_obs}
\tablecaption{Summary of radio observations and best-fit spectral measurements}
\tablewidth{0pt}
\tablehead{
\colhead{Instrument/Configuration} &
\colhead{Program ID} &
\colhead{MJD\tablenotemark{a}} &
\colhead{$t_{\rm exp}$\tablenotemark{b}} &
\colhead{$\alpha$} &
\colhead{$f_{\rm R}$(6 GHz)} &
\colhead{$\log_{10}L_{\rm R}$(5 GHz)}
\\
\colhead{} &
\colhead{} &
\colhead{} &
\colhead{(min)} &
\colhead{} &
\colhead{(mJy)} &
\colhead{(erg s$^{-1}$)}
}
\decimalcolnumbers

\startdata
 VLA/D &  18A-277 & 58398.04 & 11 & $0.230\pm0.004$ & $16.99\pm0.03$ & $30.0\pm0.1$\\ 
 VLA/D &  18A-277 & 58399.99 & 4 & $0.27\pm0.01$ & $7.46\pm0.05$ & $29.6\pm0.1$\\ 
 VLA/D &  18A-277 & 58402.85 & 4 & $0.42\pm0.02$ & $5.12\pm0.03$ & $29.5\pm0.1$\\ 
 VLA/D &  18A-277 & 58403.91 & 1 & $0.51\pm0.04$ & $4.20\pm0.04$ & $29.4\pm0.1$\\ 
 VLA/D &  18A-277 & 58405.90 & 1 & $0.42\pm0.05$ & $3.59\pm0.05$ & $29.3\pm0.1$\\ 
 VLA/D &  18A-277 & 58409.83 & 1 & $0.23\pm0.05$ & $3.19\pm0.05$ & $29.3\pm0.1$\\ 
 VLA/D &  18A-277 & 58418.85 & 3 & $0.28\pm0.03$ & $2.49\pm0.03$ & $29.2\pm0.1$\\ 
 VLA/D &  SK0335 & 58431.85 & 31 & $0.48\pm0.07$ & $1.38\pm0.07$ & $28.9\pm0.1$\\ 
 VLA/C &  SJ0238 & 58440.90 & 36 & $0.29\pm0.05$ & $1.16\pm0.01$ & $28.8\pm0.1$\\ 
 VLA/C &  18A-277 & 58473.68 & 19 & $0.50\pm0.50$ & $0.14\pm0.01$ & $27.9\pm0.2$\\ 
 VLA/C &  18A-277 & 58479.64 & 19 & $0.67\pm0.50$ & $0.15\pm0.01$ & $27.9\pm0.1$\\ 
 VLA/C &  18A-277 & 58484.75 & 19 & $-0.32\pm0.32$ & $0.15\pm0.01$ & $28.0\pm0.2$\\ 
 VLA/C &  SK0335 & 58516.70 & 38 & $-0.12\pm0.17$ & $0.156\pm0.004$ & $28.0\pm0.2$\\ 
 VLA/B &  SK0114 & 58603.32 & 36 & $-0.11\pm0.04$ & $0.88\pm0.01$ & $28.8\pm0.1$\\ 
 VLA/B &  SK0114 & 58608.35 & 36 & $0.30\pm0.07$ & $0.437\pm0.004$ & $28.4\pm0.1$\\ 
 VLA/B &  SK0114 & 58618.34 & 83 & $0.55\pm0.18$ & $0.108\pm0.003$ & $27.8\pm0.1$\\ 
 VLA/B &  SK0114 & 58627.34 & 83 & $0.44\pm0.18$ & $0.115\pm0.003$ & $27.8\pm0.1$\\ 
 VLA/B &  SK0114 & 58636.26 & 83 & $0.56\pm0.27$ & $0.076\pm0.003$ & $27.6\pm0.1$\\ 
 VLA/B &  SK0114 & 58645.24 & 83 & $1.06\pm0.34$ & $0.078\pm0.004$ & $27.6\pm0.1$\\ 
\enddata
\tablenotetext{a}{MJD at the start of the observation}
\tablenotetext{b}{On-source time, not removing time lost to flagging.}

\end{deluxetable*}

\end{document}